\documentclass[a4paper,10pt]{article}

\usepackage[margin=1in]{geometry}

\pdfoutput=1

\usepackage{graphicx}
\usepackage{bm}
\usepackage{dcolumn}
\usepackage{setspace}
\usepackage{abstract}
\usepackage{multirow}
\usepackage{sidecap}
\usepackage{caption}
\usepackage{subcaption}

\usepackage[affil-it]{authblk}

\usepackage[utf8]{inputenc}
\usepackage{lineno,hyperref}

\begin{document}
\title{ \bf Fatigue Deformation of Polycrystalline Cu Using Molecular Dynamics Simulations
}

 \date{}

\author{ G. Sainath*, P. Rohith, and B.K. Choudhary** }

\affil {Deformation and Damage Modeling Section, Mechanical Metallurgy Division \\ 
Indira Gandhi Centre for Atomic Research, Kalpakkam - 603102, Tamil Nadu, India}

\maketitle
\doublespacing
\begin{onecolabstract}

Molecular dynamics (MD) simulations have been performed to investigate the fatigue deformation behaviour 
of polycrystalline Cu with grain size of 5.4 nm. The samples were prepared using Voronoi algorithm with
random grain orientations. Fatigue simulations were carried out by employing fully reversed, total strain 
controlled cyclic loading at strain amplitude of $\pm4$\% for 10 cycles. The MD simulation results 
indicated that the deformation behaviour under cyclic loading is dominated by the slip of partial 
dislocations enclosing the stacking faults. At higher number of cycles, the grain boundary migration 
leading to coarsening of larger grains at the expense of the smaller grains has been observed. The 
cyclic stress-strain behaviour, the deformation mechanisms and the variation of dislocation density 
as a function of cyclic deformation have been discussed. \\
 
\noindent {\bf Keywords: } Molecular dynamics simulations, Polycrystalline Cu, Fatigue, Partial dislocations 
and grain growth.


\end{onecolabstract}

\renewcommand{\thefootnote}{\fnsymbol{footnote}} \footnotetext{* email : sg@igcar.gov.in}
\renewcommand{\thefootnote}{\fnsymbol{footnote}} \footnotetext{** email : bkc@igcar.gov.in}

{\small

\section{Introduction}

Fatigue deformation in polycrystalline materials is a multi-scale problem involving the crack nucleation 
at the atomic scale to the final failure at the engineering scale \cite{Review}. Therefore, understanding 
the fatigue deformation at multiple length scales becomes important in order to design the optimum microstructure 
against the fatigue failure. Due to difficulties in performing experiments at the atomic scale, most of the 
studies in the past have concentrated on the damage at the micro and/or macro scales. With the rapid advancement 
of computational capability and the availability of reliable inter-atomic potentials, molecular dynamics (MD) 
simulations have become a major tool to examine the mechanical behaviour of materials at the atomic scale. 
However, the grain sizes accessible by the MD simulations are in the order of nanometers (nm). Therefore, the 
nanocrystalline materials with a grain size in the order of few tens of nm can only be studied using MD
simulations.

Several MD simulation studies on the deformation of nanowires/nanocrystalline materials under monotonic loading 
conditions have been performed \cite{Cai,Rupert,Sainath}. Only few studies exist in the literature pertaining 
to the MD simulations on cyclic deformation of metals \cite{Rupert-Schuh,Panzarino,Schiotz}. Further, due to 
time and length scale limitations involved 
in MD simulations, the cyclic deformation studies have been carried out only for few cycles. Rupert and Schuh 
\cite{Rupert-Schuh} performed MD simulations to understand the cyclic deformation behaviour of polycrystalline Ni 
with grain sizes of 3, 4, 5, and 10 nm. The observed strengthening during cyclic deformation in Ni has been 
attributed to the grain boundary relaxation and the formation of low energy boundaries \cite{Rupert-Schuh}. Recently,
Panzarino et al. \cite{Panzarino} characterized the grain structure evolution during the cyclic deformation of 
polycrystalline Al having grain size of 5 nm for 10 cycles. The cyclic strengthening was associated with the grain 
rotation, grain growth and the formation of many twin boundaries \cite{Panzarino}. Like polycrystalline Al and Ni, 
MD simulations on the polycrystalline Cu with 5.5 nm grain size revealed the occurrence of grain coarsening during 
the cyclic deformation \cite{Schiotz}. However, it is not clear whether cyclic deformation leads to softening or, 
hardening in Cu \cite{Schiotz}. In order to understand the effect of cyclic loading on the strength and deformation 
behaviour of polycrystalline Cu, MD simulations have been performed on Cu with a grain size of 5.4 nm for 10 cycles. 
The cyclic stress-strain behaviour along with the observed deformation mechanisms with respect to number of cycles 
has been presented.

\section{Simulation Details}

MD simulations have been carried out in large scale atomic/molecular massively parallel simulator (LAMMPS) package 
\cite{LAMMPS}. MD simulations require an appropriate interatomic potential capable of modeling the plastic deformation. 
In view of this, embedded atom method potential for Cu given by Mishin et al. \cite{potential} has been used to describe 
the inter-atomic forces between the Cu atoms. This potential is widely used in the literature for studying the plastic 
deformation in Cu \cite{Sainath}. The polycrystalline Cu samples were prepared using Voronoi algorithm with random grain 
orientations \cite{Atomeye,Voronizer}. The overall size of the sample was about 10.8 $\times$ 10.8 $\times$ 21.6 nm 
(aspect ratio of 2:1) consisting of 204,000 atoms (Fig. \ref{Initial}a). 
The sample contains 32 grains with an average grain diameter of 5.4 nm. In order to examine the size effect, MD 
simulations were also performed for grain sizes 7.2 and 9.4 nm. Since the surfaces play an important role under 
cyclic loading, periodic boundary conditions were used only along the length direction and the remaining two directions 
were kept free to mimic the free surfaces. In order to relax the internal stresses, the model system was equilibrated 
by heating to 500 K and then cooling to the required temperature of 10 K, where the actual fatigue simulations were 
performed. Upon completion of equilibrium process, the fatigue simulations were carried out at strain amplitude of $\pm4$\% 
for 10 cycles under total strain controlled cyclic loading and by employing fully reversed sinusoidal waveform with a 
time period of 100 ps (Fig. \ref{Initial}b). This provided a constant strain rate of 1.6 $\times$ $10^9$ s$^{-1}$. The 
average stress is calculated using Virial expression \cite{Virial}. The visualization of atomic 
configurations is accomplished in AtomEye \cite{Atomeye} with common neighbour analysis (CNA) coloring \cite{CNA}.

\begin{figure}[h]
\centering
\includegraphics[width=10cm]{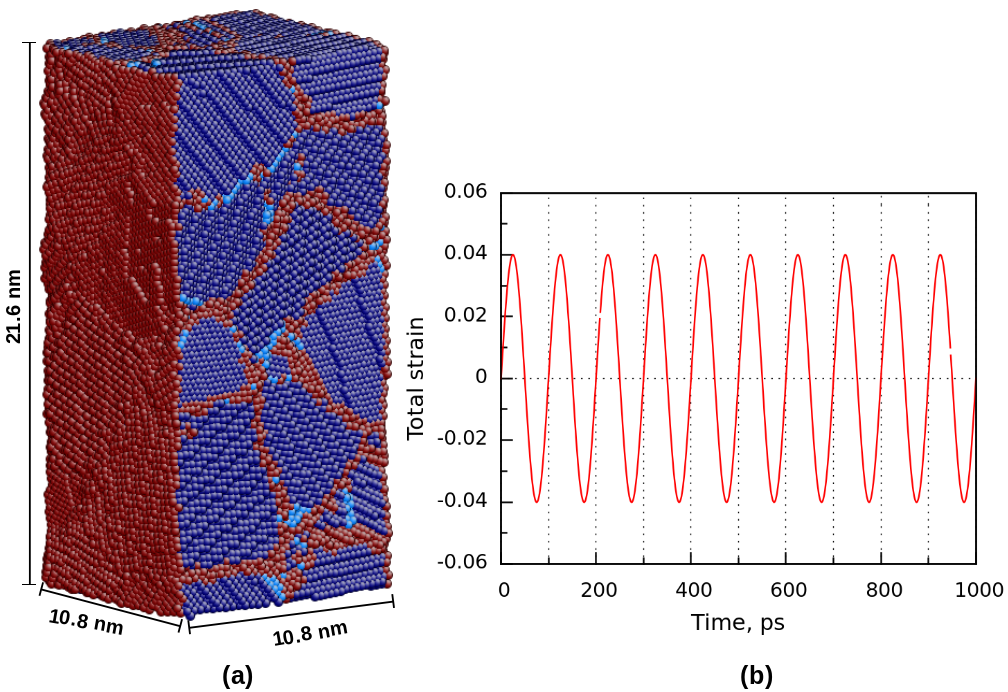}
\caption {\footnotesize (a) A typical polycrystalline Cu sample after the equilibration and (b) fatigue waveform employed 
in MD simulations are shown. The front surface was removed for clarity in (a). The atoms were colored according to the CNA 
\cite{CNA}. The red atoms represents surface and grain boundaries, blue atoms represents perfect FCC atoms and cyan atoms
represents HCP atoms. Each 100 ps equivalent to one cycle is shown as one vertical grid in (b).}
\label{Initial}
\end{figure}

\section{Results and Discussion}

Nanocrystalline materials generally exhibit wide elastic strain limits that vary with grain size, temperature 
and strain rate \cite{Rupert}. In order to identify the strain range where the plastic deformation initiates, 
MD simulations were performed on tensile deformation of polycrystalline Cu with grain size of 5.4 nm at 10 K 
(Fig. \ref{Initial}a). It has been observed that the polycrystalline Cu undergoes an elastic deformation 
up to a strain level of 0.03 followed by the irreversible plastic deformation. Based on this, the strain amplitude 
of $\pm4$\%  has been chosen for fatigue simulations. Typical stress-strain hysteresis loops corresponding to 1st, 
5th and 10th cycles are shown in Fig. \ref{stress-strain}. Similar to monotonic loading, the stress increases 
linearly during elastic deformation followed by yielding and plastic deformation up to 4\% strain. Upon reversal 
at 4\% strain the stress-strain varies linearly followed by yielding at relatively lower stress and plastic 
deformation up to -4\% strain in compression. A well defined hysteresis loop develops due to irreversible plastic 
deformation and subsequent cyclic loading resulted in cyclic hardening. Figure \ref{strength-cycles} shows the 
variations in peak cyclic stress response (CSR) in tension and compression as a function of number of cycles. 
Cyclic softening up to 3 cycles followed by continuous hardening till the end of 10th cycle can be seen in Fig. 
\ref{strength-cycles}. MD simulation results for the grain size 7.2 and 9.4 nm also exhibited similar cyclic 
stress response. The observed cyclic stress response is in agreement with those reported for nanocrystalline Ni 
\cite{Rupert-Schuh,Moser} and Al \cite{Panzarino} examined using experiments \cite{Moser} and MD simulations 
\cite{Rupert-Schuh,Panzarino}.

 \begin{figure}[h]
\centering

\begin{subfigure}[b]{0.4\textwidth}
\includegraphics[width=\textwidth]{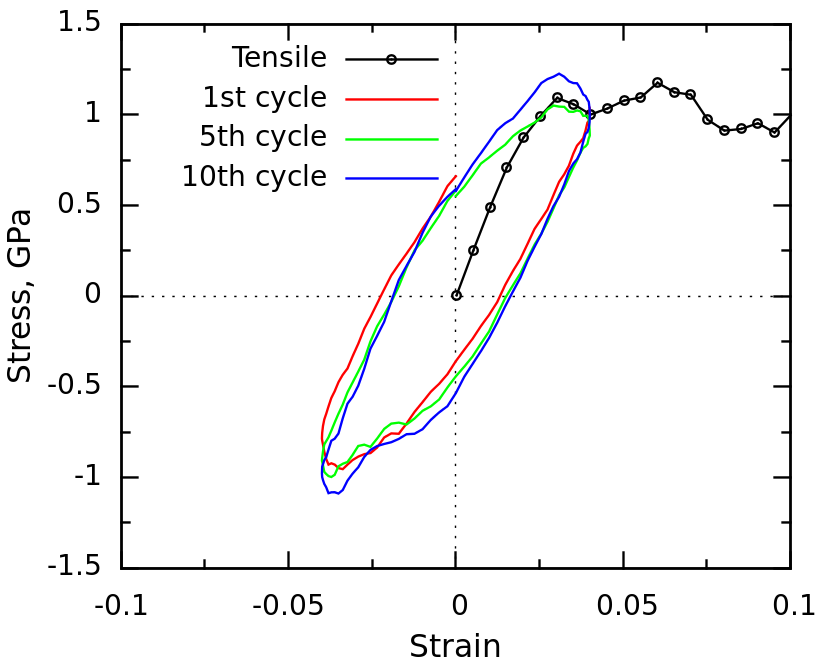}
\caption{}
\label{stress-strain}
\end{subfigure}
\qquad
\begin{subfigure}[b]{0.4\textwidth}
\includegraphics[width=\textwidth]{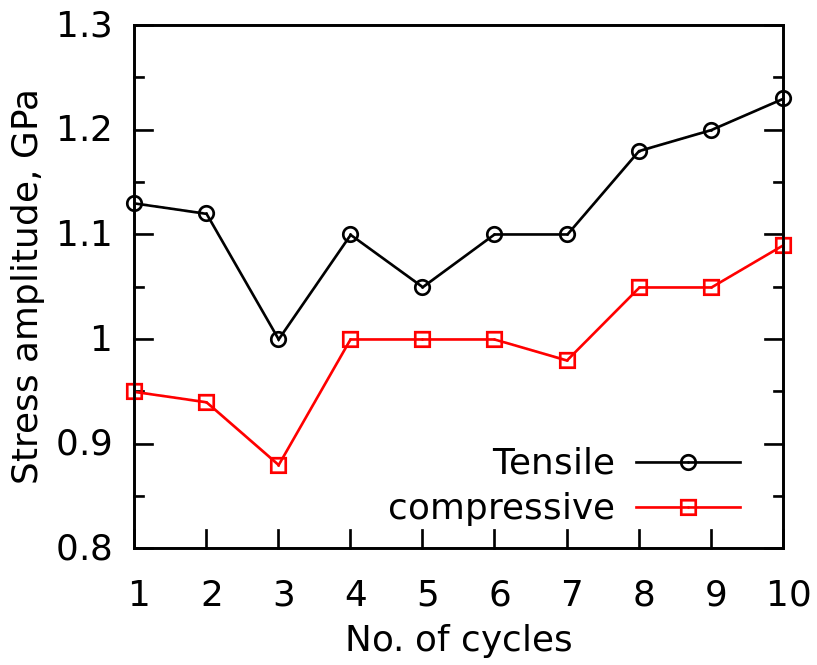}
\caption{}
\label{strength-cycles}
\end{subfigure}

\caption {\footnotesize (a) The monotonic and the cyclic stress-strain behaviour at a strain amplitude of $\pm$4\% for 
polycrystalline Cu. The peak cyclic stress response in tension and compression with number of cycles is shown in (b). } 
\label{stress-strain & strength-cycles}
\end{figure}

The progressive plastic deformation with number of cycles at 3\% total strain in tension for 1st, 5th and 10th
cycles is shown in Fig. \ref{deform}. The plastic deformation is mainly dominated by the glide of 1/6$<$112$>$ 
partial dislocations nucleating from the grain boundaries. The 1/6$<$112$>$ partial dislocations have been found 
by enclosing the stacking faults. It can be seen that, during the first cycle, the plastic deformation is activated 
only in few grains (Fig. \ref{deform}a), and with increase in the number of cycles, more grains participate in 
the deformation process (Fig. \ref{deform}b,c). The partial dislocations nucleating from one grain boundary 
(Fig. \ref{deform}a) moves towards the opposite grain boundary and gets annihilated. This suggests that the grain 
boundaries acts as a source as well as sink for dislocations. The nucleation, glide and annihilation of partial 
dislocations leave a stacking fault ribbons within the grain. Many such stacking fault ribbons produced by the 
movement of partials dislocations are shown with white arrows in Fig. \ref{deform}. The continuous nucleation of 
partial dislocations from the incoherent twin boundary (ICTB) (Fig. \ref{deform}a) and their reverse motion upon 
strain reversals converts the ICTB into a coherent twin boundary (Fig. \ref{deform}b). The coherent twin boundary 
formed during first few cycles remains highly stable during further cyclic deformation (Fig. \ref{deform}c). 
Contrary to this, the high angle grain boundaries have been found to be highly unstable under the cyclic deformation. 
It has been observed that the heavily deformed grains coarsen at the expense of undeformed grains associated with 
grain boundary migration. This grain coarsening results in the appearance of large but lower number of grains at 
the end of 10th cycle (Fig. \ref{deform}c). Our preliminary analysis indicates that the grain growth is due to 
continuous impingement of partial dislocations on the grain boundary. The observed cyclic deformation induced 
grain growth is in agreement with those reported by Panzarino et al. \cite{Panzarino} in Al and Sciotz \cite{Schiotz} 
in Cu. This is also consistent with the deformation-induced grain growth in nanocrystalline materials. 

\begin{figure}[h]
\centering
\includegraphics[width=9cm]{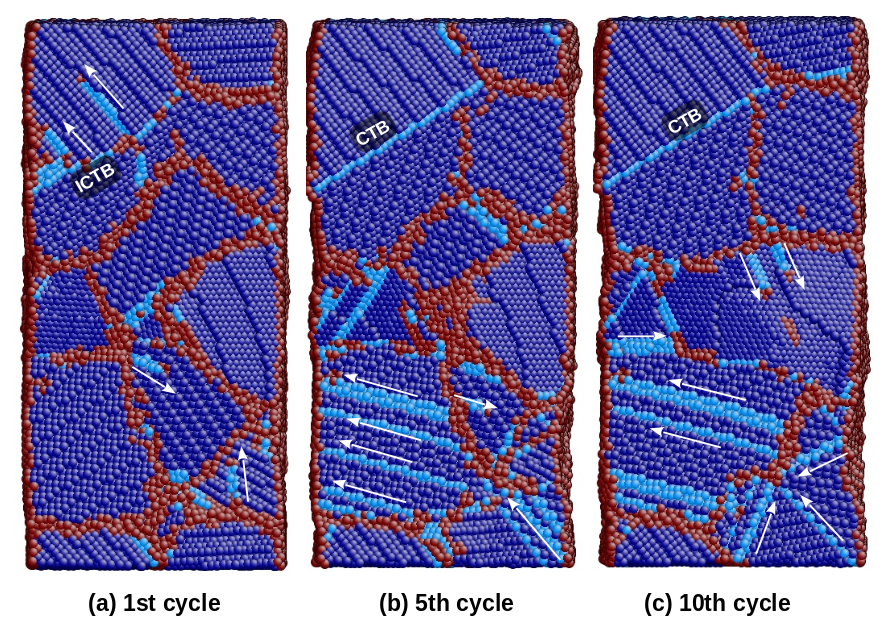}
\caption {\footnotesize The atomic snapshots showing plastic deformation in nanocrystalline Cu at a strain level 
of +3 \% for (a) 1st, (b) 5th and (c) 10th cycles. The front surface was removed for clarity. The atoms were colored
according to the CNA \cite{CNA}. The red atoms represent the grain boundaries, dislocation cores and surface atoms. 
Blue atoms represent perfect FCC atoms and cyan atoms represent HCP atoms.}
\label{deform}
\end{figure}

In order to further confirm and quantify the grain coarsening, the number of disordered atoms have been calculated 
and presented as a function of cyclic deformation in Fig. \ref{disorder}. The disordered atoms include the atoms 
in the surface, partial dislocation core and grain boundaries. By assuming the surface atoms to be constant and 
the dislocation core atoms to be low during the deformation, the observed overall decreases in \% disordered atoms 
indicates that the total grain boundary area decreases with increase in cyclic deformation, thus confirming the 
observed grain growth. The fluctuations observed in Fig. \ref{disorder} are due to fluctuations in dislocation 
core atoms during the deformation. The plastic deformation dominated by the glide of 1/6$<$112$>$ partial dislocations 
nucleating from the grain boundaries has been also observed at relatively larger grain sizes of 7.2 and 9.4 nm. 
Further, similar coarsening of heavily deformed grains at the expense of undeformed grains associated with grain 
boundary migration has been found for the grain size 7.2 and 9.4 nm.
In nanocrystalline materials, it is well known that the grain size and strength are related by the inverse Hall-Petch 
equation. Based on this, the observed increase in strength from 3rd to 10th cycles can be attributed to the increase 
in grain size \cite{Panzarino}. In order to understand the initial softening followed by hardening, the variations of 
dislocation density and \% stacking fault atoms have been calculated as a function of cyclic deformation (Fig. \ref{DD&SF}). 
The dislocation density is calculated by using dislocation extraction algorithm \cite{DXA}. The average dislocation 
density exhibited an increase up to 300 ps, i.e., 3 cycles followed by a continuous decrease up to 1000 ps or, 10th 
cycle (Fig. \ref{DD}). The initial increase in dislocation density has been associated with a similar increase in the 
\% stacking fault atoms in the first 3 cycles followed by nearly constant \% stacking fault atoms at higher number of 
cycles (Fig. \ref{Stacking-faults}). Since the deformation in all the grains is not activated in the first cycle, the 
increase dislocation density and stacking fault atoms up to 3-4 cycles is mainly due to the continuous activation of 
dislocation sources in the undeformed grains. Contrary to this, the decrease in dislocation density from 4th to 10th 
cycle arises mainly from the decreases in dislocation sources (grain boundaries) resulting from the grain growth. This 
decrease in dislocation sources leads to the hardening as observed from 3rd to the end of 10th cycle.

\begin{figure}
\centering
\includegraphics[width=6cm]{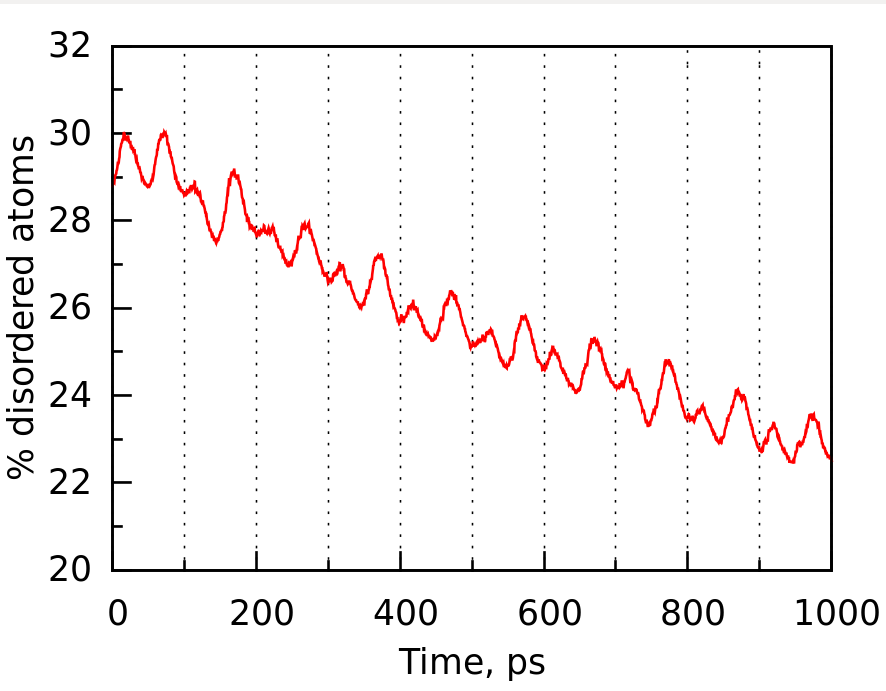}
\caption {\footnotesize The percentage of disordered atoms as a function of time during cyclic deformation. The 
disordered atoms include the grain boundary atoms, surfaces and the dislocation core atoms. Each 100 ps is equivalent 
to one cycle as shown by vertical grids.}
\label{disorder}
\end{figure}

\begin{figure}
\centering

\begin{subfigure}[b]{0.4\textwidth}
\includegraphics[width=\textwidth]{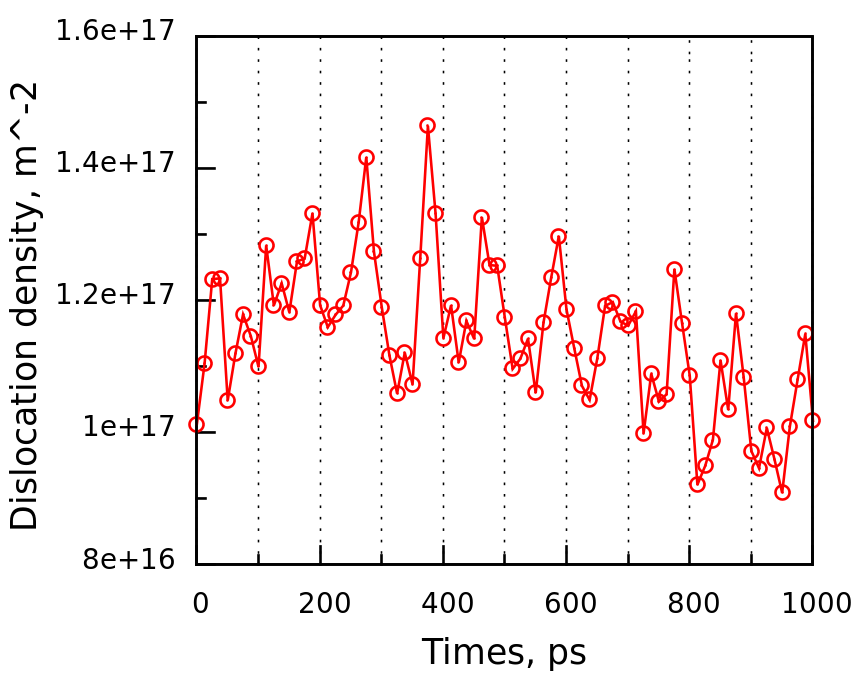}
\caption{}
\label{DD}
\end{subfigure}
\qquad
\begin{subfigure}[b]{0.39\textwidth}
\includegraphics[width=\textwidth]{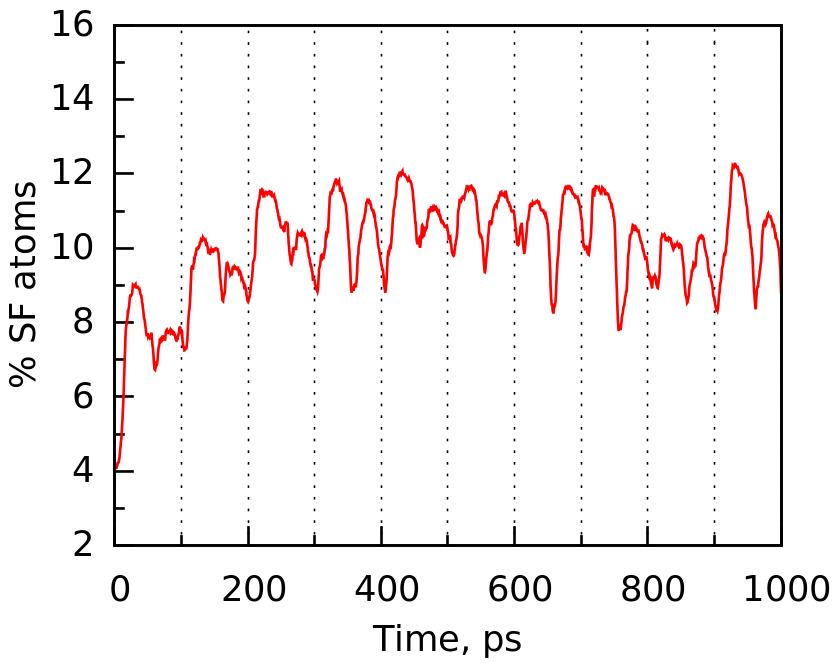}
 \caption{}
 \label{Stacking-faults}
\end{subfigure}

 \caption {\footnotesize The variations of (a) dislocation density and (b) the percentage of stacking fault atoms as a 
 function  of time during cyclic deformation of nanocrystalline Cu. Each 100 ps is equivalent to one cycle shown by 
 vertical grids.} 
 \label{DD&SF}
 \end{figure}
 
 \section{Conclusions}
 
 Large scale molecular dynamics (MD) simulations have been performed to investigate the fatigue deformation
behaviour of polycrystalline copper. The cyclic stress-strain behaviour is characterized by initial softening 
up to 3rd cycle followed by hardening till the end of 10th cycle. The MD simulation results indicated that the 
deformation behaviour under cyclic loading is dominated by the glide of partial dislocations enclosing the stacking 
faults. During the cyclic deformation, the twin boundaries were found to be stable boundaries, while the remaining 
high angle grain boundaries were highly unstable. As a result, an extensive grain growth is observed during the 
cyclic deformation. The initial softening followed by hardening has been inversely correlated with the dislocation 
density.

}

\end{document}